\RequirePackage[2020-02-02]{latexrelease}
\documentclass[aps,twocolumn,prd,superscriptaddress,preprintnumbers]{revtex4-2}

\usepackage[utf8]{inputenc}

\usepackage{mathtools}
\usepackage{amsfonts}
\usepackage{mathrsfs}
\usepackage{bbm}
\usepackage{slashed}
\usepackage{tensor}

\usepackage{graphicx}
\usepackage{array}

\usepackage{placeins}
\usepackage{makecell}
\usepackage[caption=false]{subfig}
\usepackage{float}
\usepackage{hhline}

\usepackage{xspace}
\usepackage{xfrac}
\usepackage{hyperref}
\usepackage[nameinlink]{cleveref}
\usepackage{appendix}
\usepackage{siunitx}

\usepackage{xifthen}
\usepackage{booktabs}
\usepackage{multirow}
\usepackage{courier}
\usepackage[dvipsnames]{xcolor}
\hypersetup{
	colorlinks,
	linkcolor={blue!75!black},
	citecolor={blue!75!black},
	urlcolor={blue!75!black}
}

\graphicspath{{./figures/}}

\newcommand{\tinytext}[1]{\text{\tiny{#1}}}
\newcommand{\IR}{\tinytext{IR}}
\newcommand{\UV}{\tinytext{UV}}
\newcommand{\IRUV}{\tinytext{IR/UV}}
\newcommand{\RBF}{\tinytext{RBF}}
\newcommand{\GP}{\tinytext{GP}}
\newcommand{\spec}{\tinytext{spec}}

\newcommand{\gettitle}{Non-perturbative strong coupling at timelike momenta}

\newcommand{\getHeidelbergAffiliation}{\affiliation{Institut für Theoretische Physik, Universität Heidelberg, Philosophenweg 16, D-69120 Heidelberg, Germany}}
\newcommand{\getDarmstadtAffiliation}{\affiliation{Institut für Kernphysik, Technische Universität Darmstadt, D-64289 Darmstadt, Germany}}
\newcommand{\getEMMIAffiliation}{\affiliation{ExtreMe Matter Institute EMMI, GSI, Planckstr. 1, D-64291 Darmstadt, Germany}}
\newcommand{\getMarseilleAffiliation}{\affiliation{Aix Marseille Univ, Université de Toulon, CNRS, CPT, Marseille, France}}
\newcommand{\getMITAffiliation}{\affiliation{Center for Theoretical Physics, Massachusetts Institute of Technology, Cambridge, MA 02139, USA}}
\newcommand{\getIAIFIAffiliation}{\affiliation{The NSF AI Institute for Artificial Intelligence and Fundamental Interactions}}

\hypersetup{
	pdftitle={\gettitle},
	pdfauthor={Horak, Pawlowski, Turnwald, Urban, Wink, Zafeiropoulos},
	pdfkeywords={QCD}
	    {lattice QCD}
	    {strong coupling}
		{correlation functions}
		{functional renormalisation group}
		{spectral reconstruction}
		{gaussian process regression},
	bookmarksopen=true,
	bookmarksopenlevel=2,
	bookmarksnumbered=true
}

\begin{document}

\title{\gettitle}

\author{Jan~Horak}
\getHeidelbergAffiliation

\author{Jan~M.~Pawlowski}
\getHeidelbergAffiliation
\getEMMIAffiliation

\author{Jonas~Turnwald}
\getDarmstadtAffiliation

\author{Julian~M.~Urban}
\getMITAffiliation
\getIAIFIAffiliation

\author{Nicolas~Wink}
\getDarmstadtAffiliation

\author{Savvas~Zafeiropoulos}
\getMarseilleAffiliation

\preprint{MIT-CTP/5506}

\begin{abstract}
    We compute the strong coupling constant of Landau gauge QCD in the full complex momentum plane, both directly and via spectral reconstruction. In particular, we consider the Taylor coupling given by the product of ghost and gluon dressing functions. Assuming spectral representations for the latter, we first show that also the coupling obeys such a representation. The subsequent spectral reconstruction of the coupling data, obtained from 2+1 flavour lattice QCD results for the ghost and gluon, is based on a probabilistic inversion of this representation using Gaussian process regression with analytically enforced asymptotics. In contradistinction, our direct calculation relies on earlier reconstruction results for the ghost and gluon spectral functions themselves, as well as data obtained in functional QCD. Apart from its relevance for studies of resonances or scattering processes, the calculation also serves as a non-trivial benchmark of our reconstruction approach. The results show remarkable agreement, testifying to the reliability of the method.
\end{abstract}

\maketitle

\section{Introduction}

Real-time correlation functions are of great importance in the calculation of timelike observables in quantum chromodynamics (QCD). Physical scattering processes or the hadronic resonance spectrum represent prominent examples requiring first-principle input in the form of fundamental correlation functions in Minkowski spacetime. The strong coupling constant of QCD is a central ingredient in any of those, as it describes the interaction strength between the fundamental fields. One of its most salient features is asymptotic freedom, i.e., the decay towards small distances, which is well captured by perturbation theory. In contrast, the large distance or low energy behaviour, where the coupling grows large, can only be described via non-perturbative approaches such as lattice field theory or functional methods. While these approaches have proven effective for the calculation of Euclidean correlation functions, they are not well-developed in Minkowski spacetime.

Recently, progress has been made through the \textit{spectral functional approach}~\cite{Horak:2020eng, Braun:2022mgx}, enabling a direct real-time formulation of functional methods. For applications to QCD and gravity, see~\cite{Horak:2021pfr, Horak:2022myj, Horak:2022aza, Fehre:2021eob}. On the lattice field theory side, direct real-time calculations are plagued by a severe sign problem. However, Minkowski correlation functions may also be obtained indirectly via spectral reconstruction of Euclidean data. This requires inverting the Källén-Lehmann (KL) spectral representation~\cite{Kallen:1952zz, Lehmann1954}. The applicability of Gaussian process regression (GPR) to inverse problems of this type was discussed in~\cite{10.1093/gji/ggz520}. The method has been used for the reconstruction of ghost and gluon correlators~\cite{Horak:2021syv}, the computation of glueball masses~\cite{Pawlowski:2022zhh}, and similar problems in high energy physics~\cite{DelDebbio:2021whr, Candido:2023nnb, Huang:2023ija}.

Utilising numerical reconstruction techniques to compute spectral functions has a long history in non-perturbative QCD. The underlying problem is inherently ill-conditioned, prompting the development and application of a plethora of approaches over the last couple of decades, such as the maximum entropy method~\cite{Jarrell:1996rrw, Asakawa:2000tr, Haas:2013hpa}, Bayesian inference techniques~\cite{Burnier:2013nla, Rothkopf:2016luz}, Tikhonov regularisation~\cite{Ulybyshev:2017ped, Dudal:2019gvn, Dudal:2021gif}, neural networks~\cite{Fournier_2020, Yoon_2018, Kades:2019wtd, Zhou:2021bvw, Lechien:2022ieg}, kernel ridge regression~\cite{arsenault2016projected, Offler:2019eij}, and basis expansions~\cite{Cuniberti:2001hm, Burnier:2011jq, Cyrol:2018xeq, Fei_2021, fei2021analytical}.

In this work, we establish a spectral representation for the strong coupling constant and compute its spectral function. The calculation is facilitated by the reconstruction results for the ghost and gluon spectral functions of~\cite{Horak:2021syv}, based on propagator data from 2+1 flavour lattice QCD calculations with domain wall fermions at a physical pion mass~\cite{Zafeiropoulos:2019flq, Cui:2019dwv}. In doing so, we improve on the previous reconstruction approach by incorporating known asymptotic behaviour into the GP kernel. Based on these data, we also apply GPR directly to the reconstruction of the Taylor coupling. This non-trivial benchmark of our reconstruction method yields remarkable agreement between the direct and indirect results, thereby making a strong case for the reliability of spectral reconstruction via probabilistic inversion with GPR. On the other hand, our results feature a broad range of applications in the calculation of physical observables. Knowledge of the coupling constant in the full complex plane is required, e.g., in the treatment of hadronic bound states via Bethe-Salpeter equations. Furthermore, in the calculation of physical scattering amplitudes, the strong coupling in Minkowski space is a necessary ingredient. Our main result, the spectral function of the strong coupling constant from both GPR reconstruction and via its spectral representation, is shown in \Cref{fig:spec}.

This paper is organised as follows. In \Cref{sec:spec-rep}, we derive the spectral representation for the strong coupling constant. The extension of our spectral reconstruction approach granting improved control over the asymptotics is described in \Cref{sec:GPR-reco}. Our results are presented in \Cref{sec:results} and we conclude in \Cref{sec:conclusion}.

\section{Scattering processes \& the timelike QCD coupling}\label{sec:spec-rep}

Scattering processes and decays in QCD are described in terms of $S$-matrix elements. At low energies, the operators of the physical in and out states are complicated objects in terms of the fundamental QCD degrees of freedom. For instance, a description of the Compton scattering of protons requires the definition of the proton or, more generally, the nucleon operator in terms of its partonic constituents. Since, on the fundamental level, the partons are related to quarks and gluons, the building blocks of the respective $S$-matrix elements are quark-gluon and quark-photon scattering processes.

In most partonic models the fundamental scattering processes are approximated by effective models for the exchange process, such as one-gluon exchange potentials that carry the qualitative property of the gluon mass gap in QCD in terms of an effective mass. Ideally, however, they should be constructed from tree-level processes in QCD with full propagators and vertices, both of which carry on-shell, timelike, and spacelike momenta. The final $S$-matrix is gauge-invariant, while the tree-level components making up the individual $S$-matrix element contributions are not. Moreover, the $S$-matrix admits a spectral representation, which is not necessarily present for the gauge-fixed correlation functions.

\subsection{Cross-section of quark--anti-quark scattering events and the \texorpdfstring{$S$}{S}-matrix element}

%
\begin{figure}[t]
	\centering
	\includegraphics[width=.75\linewidth]{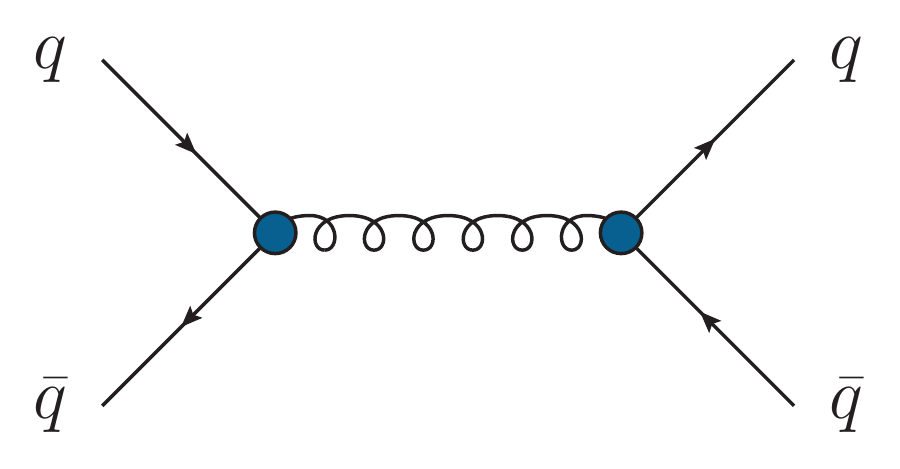}
	\caption{$q\bar q$-scattering process with a one-gluon exchange. At sufficiently large timelike exchange momenta, this process plays an important role in its respective $S$-matrix elements. Consequently, all internal quantities are dressed. Blue blobs represent full vertices, and the wiggly internal line is a full gluon propagator.}
	\label{fig:gluon-exchange}
\end{figure}

In the present work, we undertake a first step towards such a determination of non-perturbative $S$-matrix building blocks in QCD. To that end, we compute the timelike strong coupling in 2+1 flavour QCD that governs the quark--anti-quark scattering process depicted in \Cref{fig:gluon-exchange}. This diagram is at the core of many of the scattering processes used to determine the strong running coupling,
\begin{align}\label{eq:strong-coupling}
	\alpha_s(p)= \frac{g_s^2(p)}{4 \pi}
	\,.
\end{align}
It is also one of the fundamental building blocks of scattering processes in the Pomeron model~\cite{Landshoff:1986yj, Donnachie:1988nj, Ewerz:2004vf, Ewerz:2006vd}---such as the aforementioned Compton scattering of the proton---where it is typically estimated by one-gluon exchange models. For a review, see~\cite{Donnachie:2002en}; for a recent application related to the present work, see~\cite{Bopsin:2022yjn}.

Assuming the incoming and outgoing quarks $q(p)$ and anti-quarks $\bar q(\bar p)$ to be on-shell, $q\bar q$-scattering is similar to $e^+ e^-$ scattering. We expect this analogy to hold for sufficiently large timelike exchange momenta $p^2\gtrsim \SI{1}{\giga\eV\squared}$, whereas for $p^2\lesssim \SI{1}{\giga\eV\squared}$ we enter the hadronic, strongly correlated regime. There, the non-trivial embedding of the scattering quarks and anti-quarks in hadrons becomes increasingly relevant, and quark--anti-quark scattering should be also considered off-shell alongside with further, more complicated processes; for a formulation in functional approaches, see~\cite{Ewerz:2004vf}.

Here, we concentrate on the one-gluon exchange diagram as one of the building blocks of the full $S$-matrix element. The associated tree-level process shown in \Cref{fig:gluon-exchange} consists of two full quark-gluon vertices, $\Gamma^{(3)}_{q \bar q A}(p_1,p_2,p)$ with the on-shell momenta $p_1,p_2$ for the incoming as well as $\Gamma^{(3)}_{q \bar q A}(p_3,p_4,-p)$ with on-shell $-p_3,-p_4$ for the outgoing quark and anti-quark, respectively. The relative minus sign is due to the notational convention in functional computations treating all momenta as incoming. The momentum $p$ is that of the exchange gluon with the full gluon propagator $G_A(p)$. In combination, this process can be expressed as
\begin{widetext}
	\begin{align}\label{eq:Sqbarq}
		\langle q(p_3)\bar q(p_4)|\,S\,|q(p_1)\bar q(p_2)\rangle & \simeq \prod_{i=1}^4 Z^{-1/2}_q(p_i) \nonumber \\ 
		& \hspace{-2cm}\times \Biggl\{  \left[\bar u_q(p_3)\Gamma^{(3)}_{q \bar q A }(p_3,p_4,p) v_q(p_4) \right]^{a}_\mu G_A(p)\,\delta^{ab} \left(g^{\mu\nu}-\frac{p^\mu p^\nu}{p^2} \right) \left[\bar v_{q}(p_2)\Gamma^{(3)}_{q \bar q A}(p_1,p_2,-p) u_q(p_1) \right]^b_\nu\Biggr\}
		\,,
	\end{align}
\end{widetext}
where the (on-shell) quark wave functions $Z_q$ originate in the LSZ reduction formula. Note that the quark and gluon wave functions are defined such that the quark and gluon propagators $G_q(p),G_A(p)$ are proportional to $1/Z_q(p),1/Z_A(p)$, respectively. The scalar parts of the Euclidean propagators read
\begin{align}\label{eq:GAGq}
	G_{A}(p)=\frac{1}{Z_A(p)}\frac{1}{p^2}\,,\quad G_q(p) = \frac{1}{Z_q(p)} \frac{1}{p^2+M^2_q(p)}
	\,,
\end{align}
where the full propagators are proportional to the identity in color space in the adjoint (gluon) and fundamental (quark) representations. The gluon propagator in the Landau gauge also carries the projection operator on the transverse subspace (see \labelcref{eq:Sqbarq}), and the quark propagator is multiplied by $\mathrm{i}\,\slashed{p} + M_q(p)$. With \labelcref{eq:GAGq}, the standard LSZ factors carrying the pole residues are simply $Z_q^{-1/2}$, as already used in \labelcref{eq:Sqbarq}.

\begin{figure*}[t]
	\centering
	\subfloat[Spacelike Taylor coupling $\alpha_s(p)$.]{%
		\includegraphics[width=0.5\linewidth]{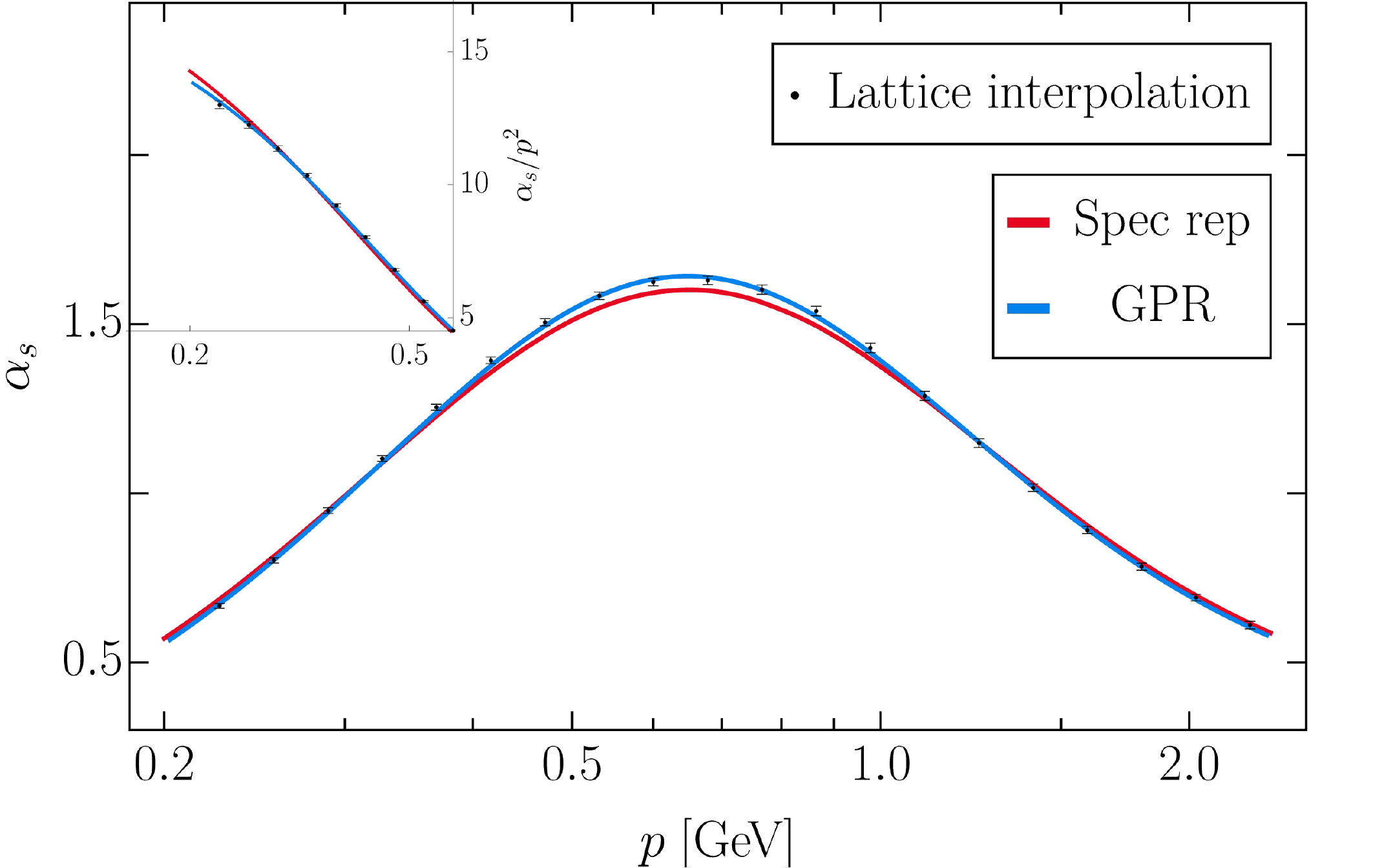}\label{fig:coupling}}%
	\subfloat[Spectral function $\rho_\alpha(\omega)$ of $\alpha_s(p)$.]{%
		\includegraphics[width=0.5\linewidth]{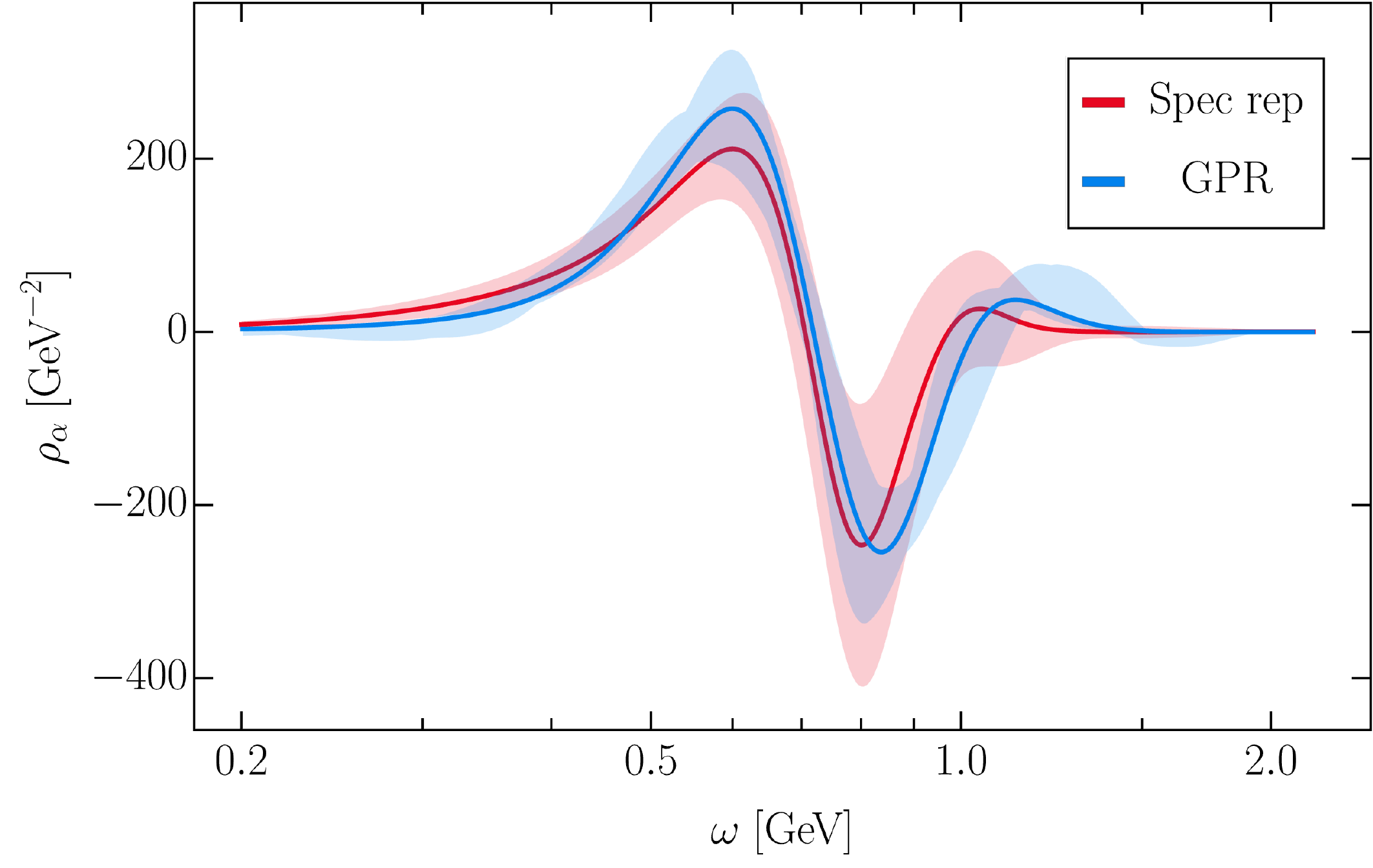}\label{fig:spec}}
	\caption{Spacelike Taylor coupling $\alpha_s$ in QCD (\Cref{fig:coupling}) and its spectral function $\rho_\alpha(\omega)$ (\Cref{fig:spec}). We compare the spectral function computed directly via~\labelcref{eq:rho_alpha_explicit} (red) to that obtained via reconstruction with GPR (blue). The direct calculation uses the reconstruction results for gluon and ghost spectral functions from~\cite{Horak:2021syv}. For the reconstruction, we use the gluon and ghost propagator data in 2+1 flavour lattice QCD from~\cite{Zafeiropoulos:2019flq, Cui:2019dwv}. Both the input spectral functions and the corresponding lattice data are displayed in \Cref{fig:input-specs}. The coupling spectral functions obtained via these two complementary approaches share all qualitative features, such as peak positions and heights as well as asymptotic behaviour. The peak structure can be connected to the respective peak structure of the gluon spectral function; see \Cref{fig:gluon-spectral}. The error band of the reconstruction result accounts for the change in the spectral function when varying the GP kernel parameters, whereas that of the direct calculation originates from propagating the uncertainty of the input. The Euclidean lattice data for the Taylor coupling $\alpha_s$ are displayed as gray squares in \Cref{fig:coupling}. We compare it to the data from its spectral representation~\labelcref{eq:coupling_spec_rep} (red) as well as the reconstruction result (blue), showing that the representation holds and that the reconstruction accurately describes the lattice data.}
	\label{fig:spec_func_rep_coupling}
\end{figure*}

The $S$-matrix element~\labelcref{eq:Sqbarq} is renormalisation group (RG) invariant, as required. To see this explicitly, we reparametrise the vertices in terms of wave functions of the legs and an RG-invariant core, 
\begin{align}\label{eq:barG3}
	\Gamma^{(3)}_{q \bar q A }(p_i,p_j,p) = Z_q^\frac12(p_i)Z_q^\frac12(p_j) Z_A^\frac12(p)\,\bar\Gamma^{(3)}_{q \bar q A }(p_i,p_j,p)
	\,,
\end{align}
where $\bar\Gamma^{(3)}_{q \bar q A}$ has the transformation properties of a running coupling, and naturally occurs in the $S$-matrix element. Inserting \labelcref{eq:barG3} into the $S$-matrix element \labelcref{eq:Sqbarq} leads us to 
\begin{widetext}
	\begin{align}\label{eq:barSqbarq}
		\langle q(p_3)\bar q(p_4)|\,S\,|q(p_1)\bar q(p_2)\rangle \simeq \left[\bar u_q(p_3)\bar\Gamma^{(3)}_{q \bar q A}(p_3,p_4,p) v_q(p_4) \right]_\mu^a  \frac{1}{p^2}\,\delta^{ab} \left(g^{\mu\nu}-\frac{p^\mu p^\nu}{p^2} \right) \left[\bar v_{q}(p_2)\bar\Gamma^{(3)}_{q \bar q A}(p_1,p_2,-p) u_q(p_1) \right]_\nu^b
		\,.
	\end{align}
\end{widetext}
We restrict ourselves to the limit of large transfer momentum $p^2 \equiv s$ of the scattering event with $p_1 p_3 = p_2 p_4 = s(1 - \cos \theta)/4$ and scattering angle $\cos \theta = \boldsymbol{p}_1 \boldsymbol{p}_3/(|\boldsymbol{p}_1||\boldsymbol{p}_3|)$. For small $s$, we approach the reliability limit of our approximations. We return to the respective discussion after deriving our results.

Additionally, in a last approximation step we concentrate on the classical tensor structure $\gamma_\mu\, T^a$ in the full quark-gluon vertex, 
\begin{align}\label{eq:alphas}
	\left[\bar\Gamma^{(3)}_{q \bar q A }(p_i,p_j,p)\right]^a_\mu\approx \mathrm{i}\,\gamma_\mu T^a \sqrt{4 \pi \alpha_s(s)}
	\,.
\end{align}
Here, $T^a$ is the $SU(3)$ generator in the fundamental representation and $\alpha_s(s)$, defined in \labelcref{eq:strong-coupling}, is the strong coupling of the quark-gluon scattering process in the $s$-channel. On the equation of motion, the $\slashed{p_i}$ terms vanish, and we obtain
\begin{align}\label{eq:S2prae}
	\left| \bar u_q(p_3)\gamma_\mu T^a  v_q(p_4) \,  \bar v_{q}(p_2)\gamma_\mu T^a u_q(p_1)\right|^2 & \nonumber \\[1ex]
	&\hspace{-2cm}\to \frac{s^2}{9}\left( 1+\cos^2 \theta\right)
	\,,
\end{align}
in the high energy limit. In \labelcref{eq:S2prae}, we have performed an average/sum over spins and color in the initial/final state. With \labelcref{eq:barSqbarq,eq:S2prae}, we arrive at 
\begin{align}\label{eq:barSqbarqSymPoint}
	\left|\langle q(p_3)\bar q(p_4)|\,S\,|q(p_1)\bar q(p_2)\rangle\right|^2 & \nonumber \\[1ex] 
	&\hspace{-1.3cm}\to \frac{1}{9}\left[ 4 \pi \alpha_s(s) \right]^2 \left( 1+\cos^2 \theta\right)
	\,.
\end{align}
with $\alpha_s(p)$ defined in \labelcref{eq:strong-coupling}. \Cref{eq:barSqbarqSymPoint} highlights the importance of the strong coupling constant $\alpha_s(s)$ for physical scattering processes. For the remainder of this work, we adopt the linear momentum argument $p = \sqrt{s}$ for the coupling.

In the present work, we shall compute the strong coupling $\alpha_s(p)$ and, hence, the above $S$-matrix element from its spectral representation for general complex frequencies, including the timelike momenta relevant for \labelcref{eq:barSqbarqSymPoint}. We utilise that the strong coupling can be computed from the quark-gluon vertex, the three- and four-gluon vertices, as well as the ghost-gluon vertex. The computation involves the wave functions $Z_q(p),Z_A(p)$ of quarks and gluons as defined in~\labelcref{eq:GAGq} and the ghost wave function $Z_c(p)$ from 
\begin{align}\label{eq:prop_dressing}
	G_{c}(p) = \frac{1}{Z_{c}(p)} \frac{1}{p^2}
	\,.
\end{align}
The avatars of the strong couplings are then defined as the (symmetric point) dressings of the classical tensor structures, see \labelcref{eq:barG3} and \labelcref{eq:alphas}.

A final definition of the strong coupling in the Landau gauge is given by the propagator or Taylor coupling, that utilises Taylor's non-renormalisation theorem for the ghost-gluon vertex. This leads to the Taylor coupling, solely defined by the ghost and gluon dressing functions, 
\begin{align}\label{eq:prop_coupl}
	\alpha_s(p) = \frac{g_s^2}{4\pi} \frac{1}{Z_A(p)Z_c^2(p)}
	\,.
\end{align}
All strong coupling avatars have the same universal two-loop running but differ for infrared momenta; see \cite{Cyrol:2017ewj}. For an evaluation of the infrared differences between the Taylor coupling and the quark-gluon coupling, see \cite{Gao:2021wun}. The latter regime is not accessible within the present approximation. Hence, we use the Taylor coupling \labelcref{eq:prop_coupl} for the evaluation of \labelcref{eq:barSqbarqSymPoint}. Its corresponding spectral function $\rho_\alpha$ is depicted in \Cref{fig:spec_func_rep_coupling}. It allows us to compute the coupling $\alpha_s(p)$ for complex frequencies including timelike momenta; see \Cref{fig:complex_plane}. Timelike result for the strong coupling in the perturbative domain can be found, e.g., in~\cite{Milton:1997us, Alekseev:2002zn}.

\subsection{Spectral representation}

For the computation of \labelcref{eq:prop_coupl}, and hence of \labelcref{eq:barSqbarqSymPoint}, we require the ghost and gluon propagators for timelike momenta. We assume that the propagators admit a KL representation, 
\begin{align}\label{eq:KL_rep}
    G(p) = \int_{-\infty} ^\infty \frac{d\lambda}{2 \pi}\, \frac{ \lambda\,\rho(\lambda) }{\lambda^2 + p^2} \equiv \int_0^\infty \mathrm{d}\lambda\, K(p,\lambda)\, \rho(\lambda)
\,,
\end{align}
where we have implicitly defined the KL kernel $K(p, \lambda)$. The spectral function $\rho$ is defined via
\begin{align}\label{eq:rho-def}
    \rho(\omega) = 2 \, \mathrm{Im} \, G\left(- \mathrm{i} (\omega + \mathrm{i} 0^+) \right)
\,,
\end{align}
with $\rho(\omega) = -\rho(-\omega)$. For the propagators of physical particles, the spectral function is the probability density for (multi-)particle excitations to be created from the vacuum in the presence of the corresponding quantum field. Consequently, in this case, the spectral function is positive semi-definite and normalisable. For propagators of `unphysical' fields, such as gauge fields, positive semi-definiteness is no longer required and the spectral representation reduces to a statement about the analytic structure of the corresponding correlation function; see, e.g.,~\cite{Cucchieri:2004mf, Lowdon:2017gpp, Lowdon:2018uzf, Cyrol:2018xeq, Bonanno:2021squ, Horak:2021pfr, Horak:2022myj}.

The ghost propagator is known to exhibit a massless particle pole in the origin, entailing a delta pole at vanishing frequency in its spectral function $\rho_c$~\cite{Horak:2021pfr}. The gluon spectral function $\rho_A$ is continuous along the whole real frequency axis and is not expected to show distributional contributions. Taking into account the explicit forms of the spectral functions, the ghost and gluon dressing functions can be expressed as
\begin{align}\label{eq:spec_rep_dressings}
    \frac{1}{Z_A(p)} = & \,  p^2 \int_0^\infty \frac{\mathrm{d}\lambda}{\pi} \frac{\lambda \, \rho_A(\lambda)}{\lambda^2 + p^2} \,, \nonumber \\[2ex]
    \frac{1}{Z_c(p)} = & \,  \frac{1}{Z_c^0} + p^2 \int_0^\infty \frac{\mathrm{d}\lambda}{\pi} \frac{\lambda \, \tilde \rho_c(\lambda)}{\lambda^2 + p^2}
\,,
\end{align}
where $1/Z_c^0$ is the residue of the massless delta pole of $\rho_c$, and $\tilde \rho_c$ denotes the continuous part.

\begin{figure*}
	\centering
	\subfloat[Ghost spectral function.]{%
		\includegraphics[width=.5\linewidth]{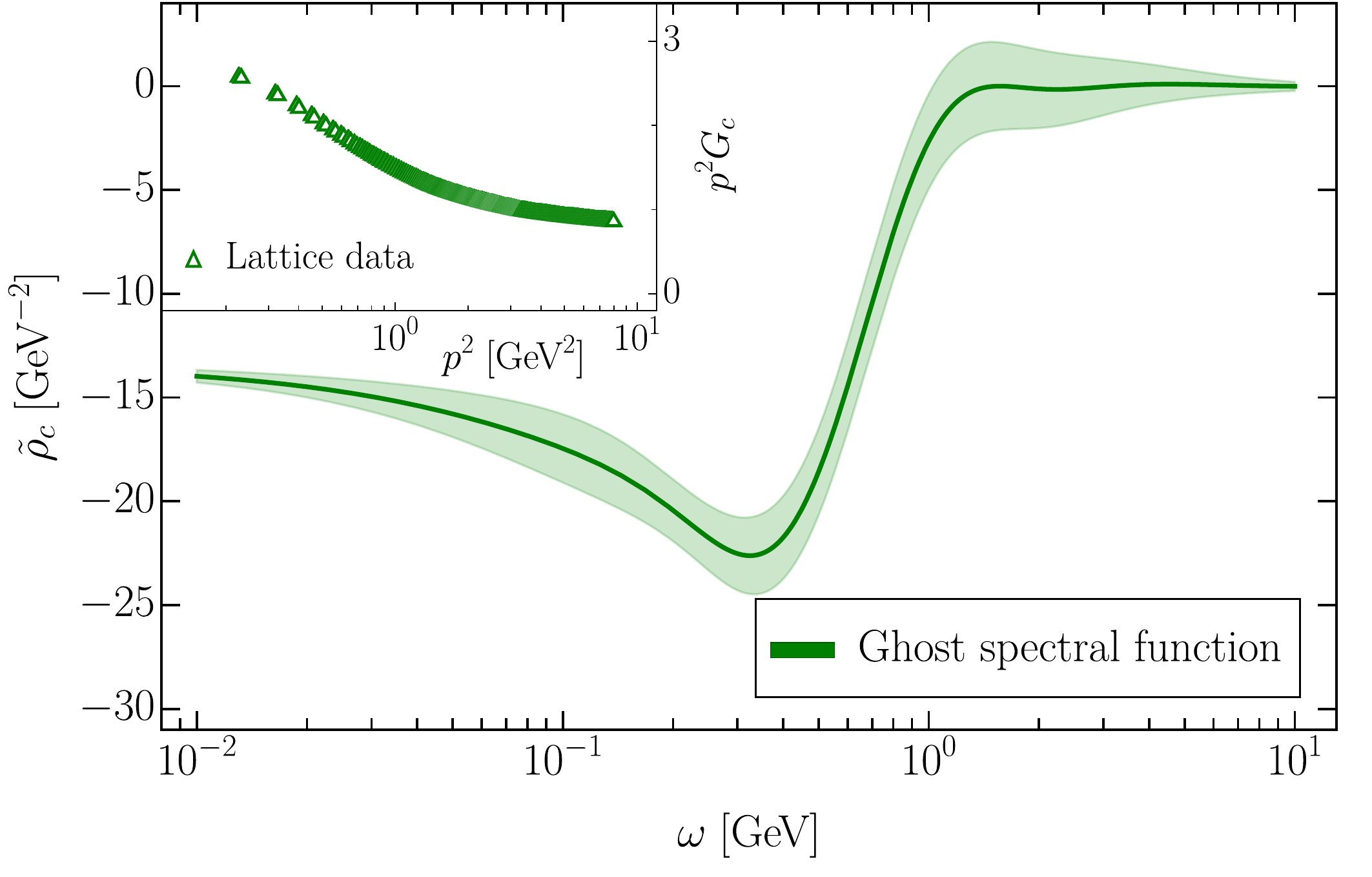}\label{fig:ghost-spectral}}%
    \subfloat[Gluon spectral function.]{%
		\includegraphics[width=.5\linewidth]{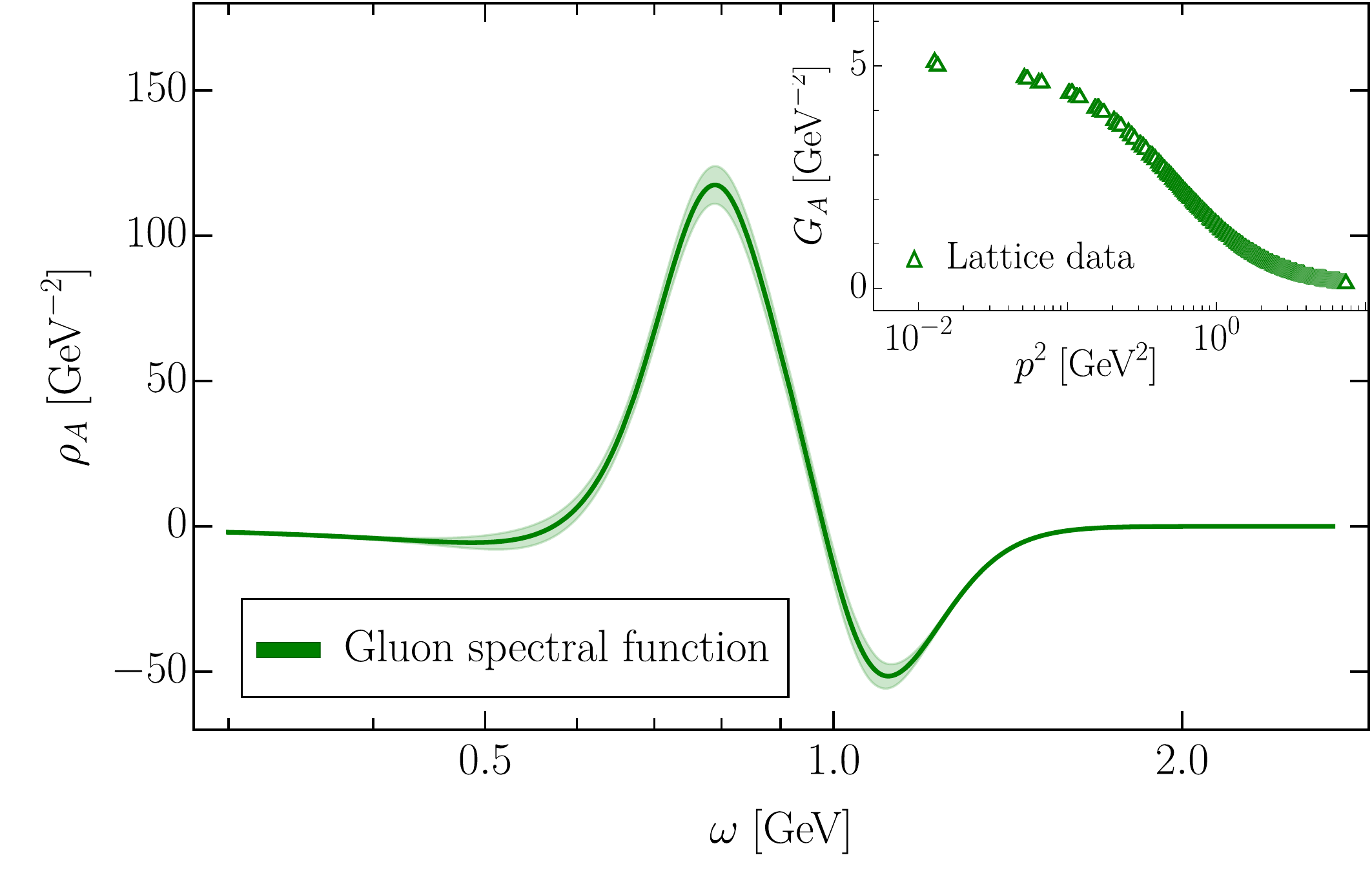}\label{fig:gluon-spectral}}%
    \caption{The continuous parts of the ghost (\Cref{fig:ghost-spectral}) and gluon (\Cref{fig:gluon-spectral}) spectral functions obtained in~\cite{Horak:2021pfr} (see also \Cref{sec:lattice-data}), used here as input for the calculation of the coupling spectral function shown in \Cref{fig:coupling} via its spectral representation~\labelcref{eq:coupling_spec_rep}. Shaded areas represent $1\sigma$-bands of the statistical error of the mean prediction based on the available observations and precision. Note that for the calculation of the gluon spectral function, the UV and IR asymptotic regimes are assumed to be maximally large. This leads to a small reconstruction error without accounting for systematics; see \Cref{app:hyperparams} for a detailed discussion.}
	\label{fig:input-specs}
\end{figure*}

Given the existence of a spectral representation, the associated correlation function must obey certain symmetries and fulfill requirements about its infrared (IR) and ultraviolet (UV) asymptotics. It can be shown that the existence of spectral representations for the ghost and gluon propagators implies the existence of such a representation also for the Taylor coupling as defined in~\labelcref{eq:prop_coupl}; see \Cref{app:spec-rep-general} for details. Specifically, it is given by
\begin{align}\label{eq:coupling_spec_rep}
    \alpha_s(p) = p^2 \int_0^\infty \mathrm{d}\lambda\, K(p,\lambda)\, \rho_\alpha(\lambda)
\,.
\end{align}
With \labelcref{eq:coupling_spec_rep}, the strong coupling spectral function is obtained from its retarded correlator via
\begin{align}\label{eq:rho_alpha}
    \rho_\alpha(\omega) = -  \frac{2}{\omega^2} \mathrm{Im} \, \alpha_s \big(-\mathrm{i} (\omega + \mathrm{i} 0^+) \big)
\,.
\end{align}
Now we use the definition of the Taylor coupling~\labelcref{eq:prop_coupl} and insert the spectral representations of ghost and gluon dressing functions~\labelcref{eq:spec_rep_dressings}. Then, the spectral function \labelcref{eq:rho_alpha} of the coupling can be written as 
\begin{align}\label{eq:rho_alpha_explicit}
    \rho_\alpha(\omega) = & \, -2 \, \mathrm{Im} \, \Bigg[ \Bigg( \int_0^\infty \frac{\mathrm{d}\lambda}{\pi} \frac{\lambda \, \rho_A(\lambda)}{\lambda^2 - \omega^2 + \mathrm{i} 0^+} \Bigg) \nonumber \\[2ex] 
    & \times \Bigg( \frac{1}{Z_c^0} - \omega^2 \int_0^\infty \frac{\mathrm{d}\lambda}{\pi} \frac{\lambda \, \tilde \rho_c(\lambda)}{\lambda^2 - \omega^2 + \mathrm{i} 0^+} \Bigg)^2 \, \Bigg]
\,.
\end{align}
Since the Taylor coupling decays logarithmically in the UV, its spectral function obeys a superconvergence condition~\cite{Horak:2021pfr, Bonanno:2021squ}, given by
\begin{align}\label{eq:superconvergence}
    \int_0^\infty \textrm{d}\lambda \, \lambda \rho_\alpha(\lambda) = 0
\,.
\end{align}
In the case of the gluon propagator, this is the well-known Oehme-Zimmermann condition~\cite{Oehme:1979bj, Oehme:1990kd}.

A treatment of the analytic low-frequency behaviour of continuous parts of the spectral functions has been initiated in~\cite{Cyrol:2018xeq}. In particular, it was shown that for correlation functions obeying a KL representation, a simple relation between the IR asymptotics of the correlator and its spectral function can be derived by differentiating with respect to the frequency. For the Taylor coupling we explicitly find
\begin{align}\label{eq:rho_alpha_IR}
    \lim_{\omega \to 0^+} \partial_\omega \rho_\alpha(\omega) = -2 \lim_{p \to 0^+} \partial_p  \frac{\alpha_s(p)}{p^2}
\,.
\end{align}
Hence, if the coupling approaches zero in the origin faster than $p^2$, we expect the spectral function to approach zero from below, and vice versa.

\subsection{Lattice data}\label{sec:lattice-data}

During the past two decades, lattice QCD results for Landau gauge two-point functions have advanced to an impressive quantitative level of precision; see, e.g.,~\cite{Bonnet:2000kw, Sternbeck:2005tk, Boucaud:2005gg, Silva:2005hb, Cucchieri:2006tf, Cucchieri:2008qm, Oliveira:2008uf, Bogolubsky:2009dc, Iritani:2009mp, Ayala:2012pb, Athenodorou:2016oyh, Boucaud:2017obn, Duarte:2016ieu, Aguilar:2019uob, Aguilar:2021lke, Aguilar:2021okw}. A recent review of lattice and functional results can be found in~\cite{Ferreira:2023fva}. The lattice ghost dressing function and gluon propagator data used in this work have been obtained from recent calculations with 2+1 dynamical fermion flavours at the physical point~\cite{Zafeiropoulos:2019flq, Cui:2019dwv}. In particular, the ensembles of gauge configurations were generated by the RBC/UKQCD collaboration in~\cite{Allton:2007hx, Allton:2008pn, Arthur:2012yc, Blum:2014tka, Boyle:2017jwu}, leveraging the Iwasaki gauge action~\cite{Iwasaki:1985we} and the domain wall fermion action~\cite{Kaplan:1992bt, Shamir:1993zy} with a pion mass of \SI{139}{\mega\eV}. This choice of action (with a particular implementation of the Möbius kernel~\cite{Brower:2004xi}) exhibits favourable chiral properties with a much smaller size in the fifth dimension than required by conventional domain wall fermions. These ensembles were utilised in~\cite{Zafeiropoulos:2019flq, Cui:2019dwv} for the calculation of the ghost and gluon propagators as well as the running of the strong coupling in the Taylor (miniMOM) scheme~\cite{Sternbeck:2007br, Boucaud:2008gn, Sternbeck:2010xu}, and an associated effective charge~\cite{Binosi:2016nme}.

The continuum limit of the lattice data is only obtained with a proper treatment of discretisation effects. For the Landau gauge propagators this is achieved by an analysis of the physical scaling violation as described in~\cite{Boucaud:2018xup}, leading to continuum extrapolated propagators with the correct momentum running. The resulting gluon propagator and ghost dressing data are displayed in the insets of \Cref{fig:input-specs}. These data combined with results from functional Yang-Mills theory and QCD~\cite{Cyrol:2016tym, Cyrol:2017ewj, Fu:2019hdw, Gao:2021wun} have also been reconstructed in~\cite{Horak:2021syv}.

Since the lattice data for the propagators are available only on different momentum grids, the coupling as defined in \labelcref{eq:prop_coupl} is computed from interpolations of the respective dressings. These are obtained by direct GPR and therefore assume no general features of the underlying correlators apart from continuity. We compute the coupling including errors for $600$ logarithmically spaced points between $\SI{0.23}{\giga\eV}$ and $\SI{2.69}{\giga\eV}$. For technical convenience, the coupling is extended perturbatively at large momenta in order to control the amplitude of the UV asymptotics; see \Cref{app:coupling-asymp}. A subset of these data is shown in \Cref{fig:coupling}. Here, we replace the error with the difference between the values computed as described above, and the coupling obtained from the product of the ghost and gluon spectral functions, described around \labelcref{eq:rho_alpha_explicit} and in \Cref{sec:results}.

\section{GPR reconstruction with controlled asymptotics}\label{sec:GPR-reco}

GPR is a popular framework for the probabilistic modelling of functions from a finite number of data points; see~\cite{kanagawa2018gaussian, liu2019gaussian} for recent reviews and~\cite{10.5555/1162254} for a textbook account. Example applications in high energy physics include the computation of parton distribution functions~\cite{Alexandrou:2020tqq} and modelling backgrounds in detectors~\cite{Frate:2017mai}. The method can be used to predict solutions to linear inverse problems~\cite{10.1093/gji/ggz520}, i.e., when the only available data are indirect observations of the desired function after a linear forward process. This makes the approach suitable for spectral reconstruction. Importantly, it does not in general require choosing a particular functional basis. This avoids many of the numerical artifacts like additional peak structures that are commonly encountered when employing reconstruction algorithms with predetermined families of solutions, due to the presence of unrepresentable features. We summarise the main concepts in \Cref{app:GPR}; see also~\cite{Horak:2021syv, Pawlowski:2022zhh} for a comprehensive introduction as well as further details and references.

As an extension to this approach, in this paper we introduce a novel technical improvement that allows us to explicitly control the asymptotic behaviour of the predictions by specifying concrete functional forms in the appropriate limits only, without restricting the expressivity of the GP model in the region of interest. When considering different design choices for GPs, one often opts for so-called \textit{universal} kernels. One prominent example also used in the present work is the radial basis function (RBF) kernel~\labelcref{eq:rbfkernel}. The basis of kernel eigenfunctions of such universal kernels is infinite-dimensional. This allows for great flexibility in the reconstruction---universal kernels can describe any continuous function~\cite{STEINWART2002768}. However, the GPR framework allows us to also incorporate further available prior information into the predictive distribution. In the context of spectral functions, the asymptotics in the IR and UV are often analytically tractable with perturbative or functional calculations, as well as formal relations to Euclidean data like~\labelcref{eq:rho_alpha_IR}. Hence, it is beneficial to introduce a bias by reducing the space of kernel eigenfunctions to the known behaviour of the target function. This can be achieved by applying Mercer's theorem~\cite{mercer1909xvi} and constructing a kernel from the known asymptotic function $\phi(\omega)$ as
\begin{align}\label{eq:asymp_kern}
    C(\omega, \omega') = \phi(\omega) \cdot \phi(\omega')
\,.
\end{align}
Since the asymptotic behaviour is only specified in the appropriate limits, the full kernel is constructed as a combination of universal and restricted kernels using smooth step functions. With this approach, it becomes possible to smoothly transition between regions with an unknown functional basis---where a generic kernel like RBF is used---and regions with a specified basis; see \Cref{app:kernel-asymp} for further details.

\begin{figure*}[t]
	\centering
	\subfloat[Real part of the Taylor coupling $\alpha_s(\omega)$.]{%
		\includegraphics[width=0.47\linewidth]{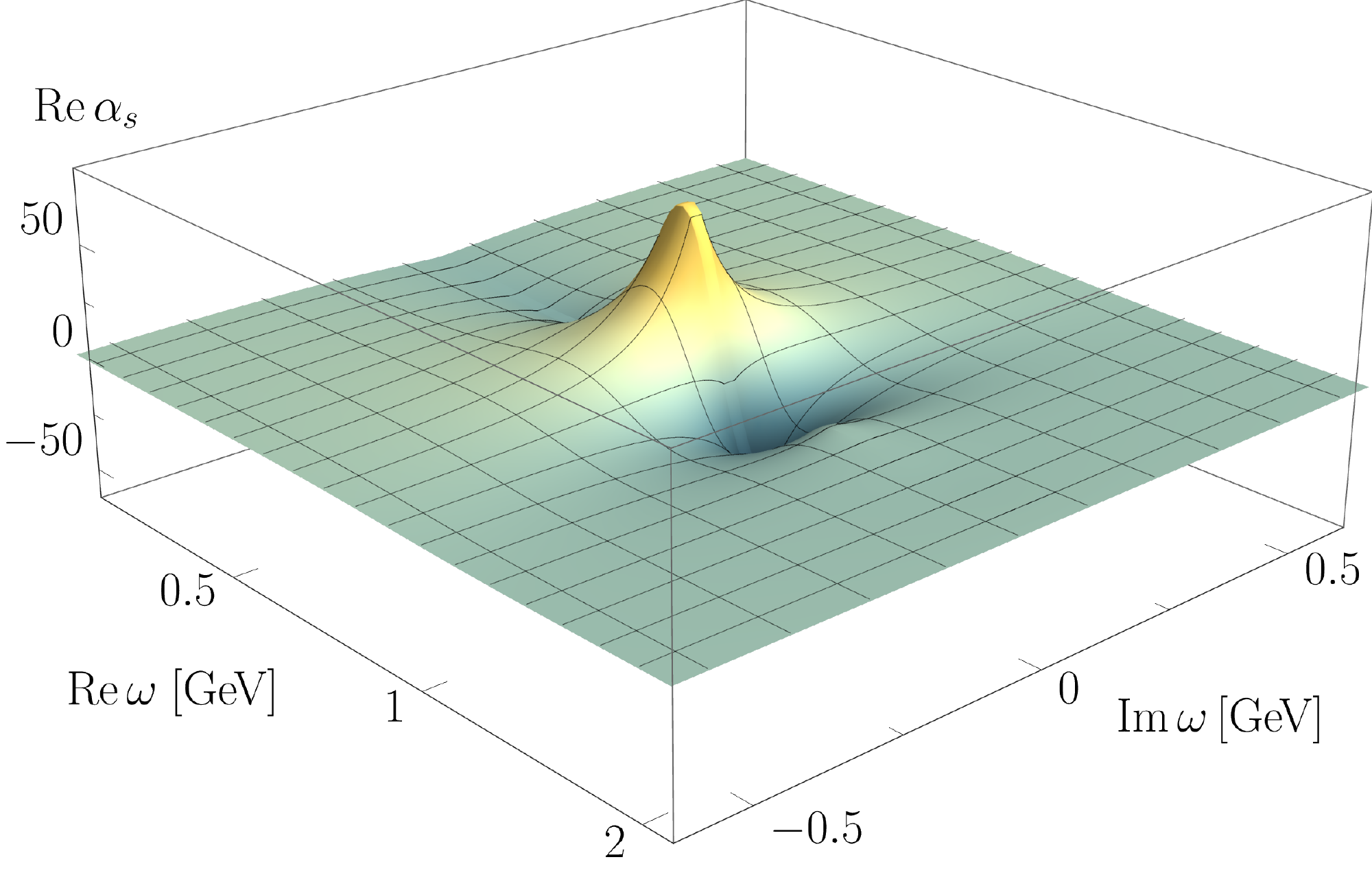}\label{fig:Realalphas}}%
	\hfill
	\subfloat[Imaginary part of the Taylor coupling $\alpha_s(\omega)$.]{%
		\includegraphics[width=0.47\linewidth]{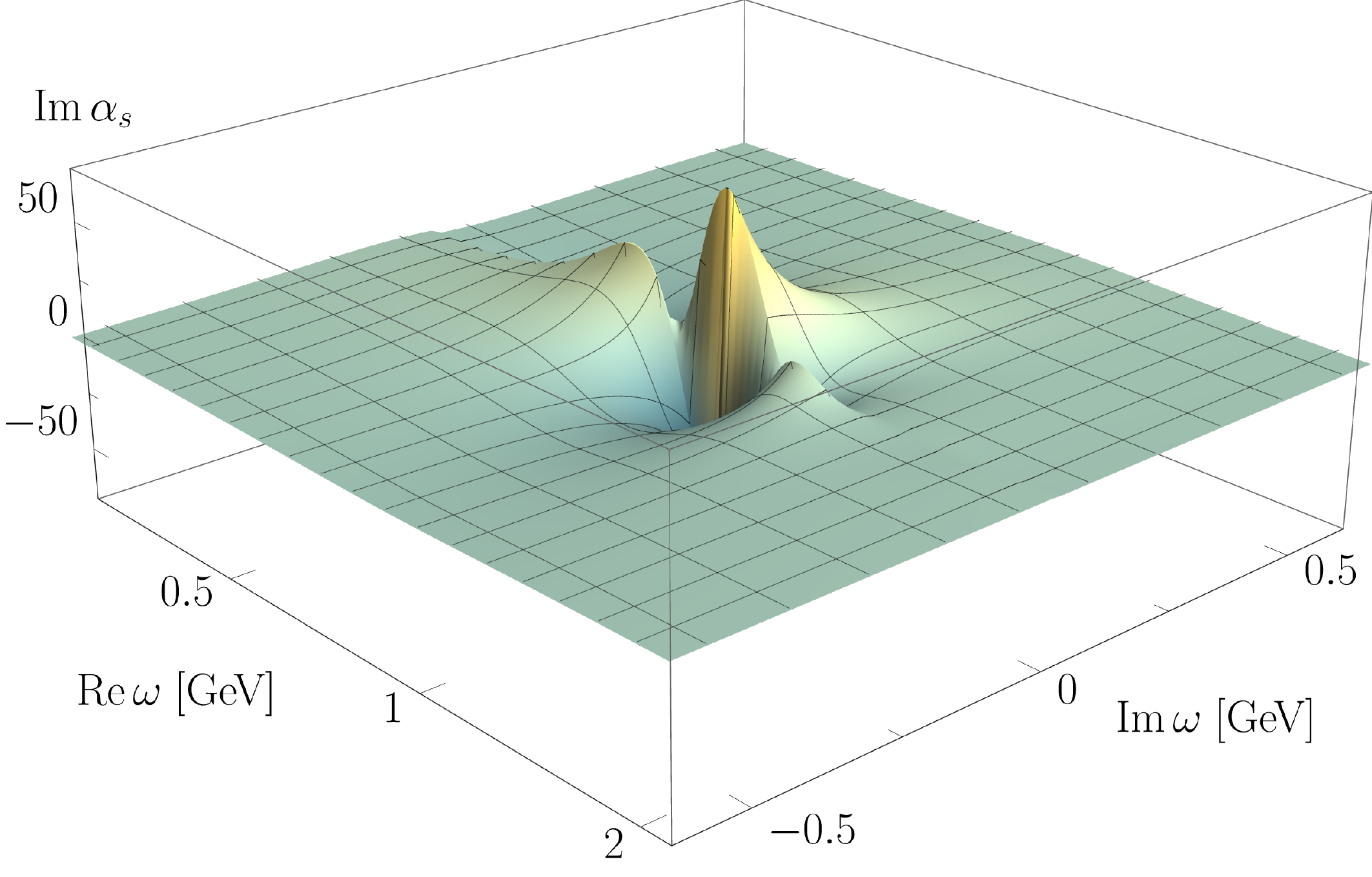}\label{fig:Imalphas}}
    \caption{Taylor coupling $\alpha_s(\omega)$ of 2+1 flavour QCD defined in~\labelcref{eq:prop_coupl} in the complex frequency right half plane (positive real frequencies), real (\Cref{fig:Realalphas}) and imaginary part (\Cref{fig:Imalphas}). The imaginary part explicitly shows the branch cut along the real frequency axis. The spectral function corresponds to the imaginary part of $\alpha_s$ at the upper half plane boundary of the branch cut, divided by $\omega^2$. Both, the real and imaginary part, exhibit distinctive peaks which can be connected to the peak structure of the gluon spectral function; see \Cref{fig:gluon-spectral}. The coupling decays logarithmically for increasing $|\omega|$.}
	\label{fig:complex_plane}
\end{figure*}

\section{Results}\label{sec:results}

Our main result, the spectral function of the Taylor coupling~\labelcref{eq:rho_alpha} in QCD, is displayed in \Cref{fig:spec}. It shows two variants: $\rho_\alpha^\GP$ from the reconstruction of the lattice QCD data via GPR, and $\rho_\alpha^\spec$ from the direct calculation based on the spectral representations of ghost and gluon propagators~\labelcref{eq:rho_alpha_explicit}. The associated input spectral functions are shown in \Cref{fig:input-specs}. In this context, we have improved the reconstruction of the gluon propagator reported in~\cite{Horak:2021syv} by explicitly incorporating the known IR and UV asymptotics with the method described in \Cref{sec:GPR-reco}. The error band of $\rho_\alpha^\spec$ is obtained by propagating the errors of these input data. Importantly, the coupling spectral functions obtained via these two different approaches agree well within errors and share all qualitative features, such as peaks and asymptotic behaviour. In both results, we can identify two prominent peaks of similar size in positive and negative direction at roughly \mbox{$\sim$\,\SI{0.6}{\giga\eV}} and \mbox{$\sim$\,\SI{0.8}{\giga\eV}}, along with a smaller positive peak at \mbox{$\sim$\,\SI{1.1}{\giga\eV}}. The spectral function $\rho_\alpha^\spec$~\labelcref{eq:rho_alpha_explicit} allows for a direct interpretation of this behaviour: it is connected to the  peak structure of the gluon spectral function, which carries information about the gluon mass gap; see \Cref{fig:gluon-spectral}. This information is extracted reliably from the lattice data with the GPR reconstruction. 

In the reconstruction of the coupling, the correct asymptotic behaviour is enforced by smooth step functions at transition points $\mu_\IR$ and $\mu_\UV$, while fully retaining the flexibility in the enclosed region where the GP kernel remains unrestricted and universal, see \Cref{app:kernel-asymp} for details. As mentioned above, this procedure has also been applied to the ghost and gluon spectral functions used here. 
It enhances significantly the stability and reliability of the prediction by connecting it to analytic results at low and high frequencies, ensuring agreement with functional and perturbative results in the relevant limits without reducing the expressivity of the GP model in the domain of interest. While the prediction shows some variation with the choice of the transition midpoints, the peak positions and heights remain remarkably stable; see \Cref{fig:aysmptotics_comparison}. Hence, we choose the size of the regions dominated by the asymptotics to be as large as possible without increasing the $\chi^2$ error of the reconstruction significantly; see \Cref{app:hyperparams} and \Cref{fig:parameterscan} for details.
Furthermore, changing the parameters controlling the transition to the asymptotic behaviour accounts for the majority of the variation in the spectral function, while changing the parameters of the RBF kernel produces errors at least one order of magnitude smaller; see \Cref{app:hyperparams} for details.
The numerical values of all kernel hyperparameters are listed in \Cref{tab:hyperparams}. Accordingly, the size of the dynamical region carrying information about the QCD mass gap is minimised, supporting the gluonic quasi-particle picture employed in various applications such as bound state studies and transport computations. Specifically, this suggests dismissing smaller negative peaks close to the dominant quasi-particle peak---they merely reflect the asymptotic behaviour and the superconvergence condition~\labelcref{eq:superconvergence}. As such, they are sensitive to changes in the gauge fixing parameter and infrared closure. This suggests that they carry physically relevant information only on a subleading level.

In \Cref{fig:coupling}, we compare the reconstructed Euclidean Taylor couplings to the result computed from the lattice data for the ghost and gluon propagators, as described in \Cref{sec:lattice-data}. Using the dressing function data obtained in this way, the resulting coupling is shown to decay towards small and large momenta. In correspondence to the scale of the peaks of the spectral function---reflecting the mass gap of the theory---also the peak of the coupling itself appears at \mbox{$\sim$\,\SI{0.6}{\giga\eV}}.

The blue curve in \Cref{fig:coupling} represents the GPR reconstruction of the Taylor coupling lattice data, corresponding to $\rho_\alpha^\GP$.
The red curve represents the coupling obtained via its spectral representation~\labelcref{eq:coupling_spec_rep} using the directly computed spectral function $\rho_\alpha^\spec$.
The calculation involves finite precision, both in the input data and in the integration.
Hence we expect a small, but not negligible, relative error.
The decent agreement between this result and the lattice/GPR reconstruction result provides a highly non-trivial benchmark check.
The error is well within our expectations, since the result obtained from the directly computed spectral function depends on the reconstructions of the gluon and ghost propagators.
If the ghost and gluon spectral functions were describing their respective propagator data to infinite precision, we would also expect perfect agreement from analytic considerations; see \Cref{app:spec-rep-general}.
Hence, the small difference can be attributed to systematic uncertainties present in the calculation.
Please note that they do not contribute to the error bands, corresponding to the purely statistical error, shown in \Cref{fig:spec_func_rep_coupling}.

In the inset of \Cref{fig:coupling}, we also show the Taylor coupling divided by $p^2$ for small Euclidean momenta $p$. The derivative of this quantity is connected to the asymptotic behaviour of the spectral function in the IR by~\labelcref{eq:rho_alpha_IR}. We observe that in the region where lattice data are available, the slope of $\alpha_s/p^2$ is negative. In accordance with the analytic requirement~\labelcref{eq:rho_alpha_IR}, the slope of the spectral function is observed to be positive in this regime.

Finally, in \Cref{fig:complex_plane} we display the real and imaginary parts of the coupling in the full complex momentum plane.
The data are obtained by evaluating the coupling spectral representation~\labelcref{eq:coupling_spec_rep} with the directly calculated spectral function $\rho_{\alpha}^\spec$ in the complex plane.
The branch cut in the imaginary part, responsible for the spectral representation, is clearly visible. As expected, no further non-analyticities in the complex plane are encountered and the coupling shows the expected decay behaviour towards large frequencies.

\section{Conclusion}\label{sec:conclusion}

In this work, we have presented results for the spectral function of the strong coupling constant in QCD obtained through a direct calculation as well as a reconstruction via GPR. Assuming spectral representations for the ghost and gluon, we have derived the spectral representation of the Taylor coupling, which is fully determined by the ghost and gluon dressing functions. With this relation, we have calculated the associated spectral function as well as the coupling itself in the full complex plane; see \Cref{fig:spec_func_rep_coupling,fig:complex_plane}. The required ghost and gluon spectral functions have been obtained using the same reconstruction method, explicitly taking into account the known asymptotic IR and UV behaviour; see \Cref{fig:input-specs}. This is facilitated by expanding the GP kernel in suitable eigenfunctions based on Mercer's theorem, extending the algorithm previously applied in~\cite{Horak:2021syv, Pawlowski:2022zhh}. The modification substantially improves the reliability of the approach by properly encoding the analytically tractable regimes into the prediction while preserving the expressivity and universality of the GP model in the region of interest. This extension of GPR is completely generic and may also be useful in other contexts where some analytic properties of a function to be modelled are known a priori, in particular if data scarcity is an issue.

A comparison of the results from the direct calculation and GPR reconstruction shows excellent agreement between both approaches; see \Cref{fig:spec}. This independent verification provides strong support for the accuracy of the computed spectral function and also underlines the power of probabilistic inversion with GPR as a spectral reconstruction approach. In particular, the findings demonstrate that uncertainty estimates obtained within this framework are reasonable, allowing to reliably quantify the expected errors in potential downstream applications based on reconstruction results. In future work, we plan to also analyse the quark-gluon vertex coupling directly based on available lattice data; see, e.g.,~\cite{Kizilersu:2021jen}.

Our results find direct application in the calculation of non-perturbative, physical scattering processes, where the strong coupling constant needs to be known at timelike momenta. While neglecting angular dependencies, the Taylor coupling considered here carries the correct RG running and hence scale-dependence of the strong coupling constant. Furthermore, it encodes genuine non-perturbative information through the input ghost and gluon dressing functions obtained from 2+1 flavour lattice QCD. Our work hence paves the way for incorporating non-perturbative information from lattice field theory to functional methods in the calculation of timelike scattering processes.

\begin{acknowledgements}
We thank J.~Papavassiliou and J.~Rodr\'iguez-Quintero for discussions and collaboration in the early stages of this work. We also thank P.~Boucaud and F.~De~Soto for an earlier collaboration and for their help in the preparation of the lattice data. We are indebted to the RBC/UKQCD collaboration, especially to P.~Boyle, N.~Christ, Z.~Dong, N.~Garron, C.~Jung, B.~Mawhinney, and O.~Witzel, for access to the lattices used in this work. Numerical computations have used resources of CINES, GENCI IDRIS (project id 52271), and of the IN2P3 computing facility in France. This work is funded by the Deutsche Forschungsgemeinschaft (DFG, German Research Foundation) under Germany’s Excellence Strategy EXC 2181/1 - 390900948 (the Heidelberg STRUCTURES Excellence Cluster) and the Collaborative Research Centre SFB 1225 (ISOQUANT). JT and NW acknowledge support by the Deutsche Forschungsgemeinschaft (DFG, German Research Foundation) – Project number 315477589 – TRR 211. NW acknowledges the support by the State of Hesse within the Research Cluster ELEMENTS (Project ID 500/10.006). JMU is supported in part by Simons Foundation grant 994314 (Simons Collaboration on Confinement and QCD Strings) and the U.S.\ Department of Energy, Office of Science, Office of Nuclear Physics, under grant Contract Number DE-SC0011090. This work is funded by the U.S.\ National Science Foundation under Cooperative Agreement PHY-2019786 (The NSF AI Institute for Artificial Intelligence and Fundamental Interactions, \url{http://iaifi.org/}). JH and JMU acknowledge support by the Studienstiftung des deutschen Volkes.
\end{acknowledgements}

\appendix

\section{Spectral representation}\label{app:spec-rep-general}

Any product of correlation functions obeying a spectral representation allows for such a representation itself. This follows from a set of sufficient conditions for the existence of a spectral representation for an arbitrary correlation function $\mathcal{C}$:
\begin{enumerate}
	\item[(i)] \textit{Holomorphicity:} $\mathcal{C}$ is holomorphic in the upper half plane ${\mathbb{H} = \{ z \, \vert \,  \mathrm{Im} \, z > 0 \} }$;
	\item[(ii)] \textit{Mirror symmetry:} $\mathcal{C}(z) = \bar{\mathcal{C}}(\bar z)$ and $\mathrm{Im} \, \mathcal{C}(z) = 0$ for $\mathrm{Im} \, z = 0$, $\mathrm{Re} \, z > 0$;
	\item[(iii)] \textit{Asymptotic decay:} $\vert z \, \mathcal{C}(z) \vert \to 0$ for $\vert z \vert \to \infty$;
	\item[(iv)] \textit{Spectral convergence:} 
		\begin{align}
			\text{(IR)} \quad & \vert z \, \text{Im} \, \mathcal{C}(z) \vert < \infty \quad \text{for} \quad z \to 0 \,, \nonumber \\
			\text{(UV)} \quad & \vert \log z \, \text{Im} \, \mathcal{C}(z) \vert \to 0 \quad \text{for} \quad z \to - \infty \nonumber
		\,.
		\end{align}
\end{enumerate}
Heuristically, (i) and (ii) guarantee that the spectral kernel has the form $1/(z + \lambda^2)$ and the spectral function is defined via~\labelcref{eq:rho-def}. The integration domain is restricted to $\lambda^2 > 0$ by (iii). Condition (iv) guarantees the convergence of the spectral integral.

It is immediately clear that for any two correlation functions $\mathcal{C}_1$, $\mathcal{C}_2$ satisfying (i)-(iii), their product $\mathcal{C} = \mathcal{C}_1 \mathcal{C}_2$ does as well. Similarly, this also applies to (iv) (UV), stating that the spectral function $\rho \sim \text{Im} \, \mathcal{C}$ decays fast enough for the spectral integral to converge in the UV, due to (iii). The infrared convergence condition (iv) (IR) does not need to be fulfilled; consider, e.g.,~$\mathcal{C}_1 = \mathcal{C}_2 = (1/z)^{\alpha}$ with $1/2 < \alpha < 1$. Nevertheless, this can be always remedied by multiplying with an appropriate power of $z$. Note that this does not violate the other conditions.

The spectral representation for the strong coupling constant is then constructed as follows: by the assumption of ghost and gluon propagator obeying the KL representation, their dressing functions obey (i) and (ii). Since by its definition~\labelcref{eq:prop_coupl} the coupling is dimensionless, (iii) does not hold. However, division by $p^2$ makes (iii) and (iv) hold true. Hence, a KL representation for $\alpha_s(p)/p^2$ is constructed. Multiplying this representation by $p^2$, we obtain the spectral representation for $\alpha_s$~\labelcref{eq:coupling_spec_rep}.

\section{Asymptotic behaviour of the strong coupling}\label{app:coupling-asymp}

\begin{figure*}[t]
	\centering
	\subfloat[Scan of $\rho_\alpha$ along the flat $\mu_\IR$ direction]{%
		\includegraphics[width=.5\linewidth]{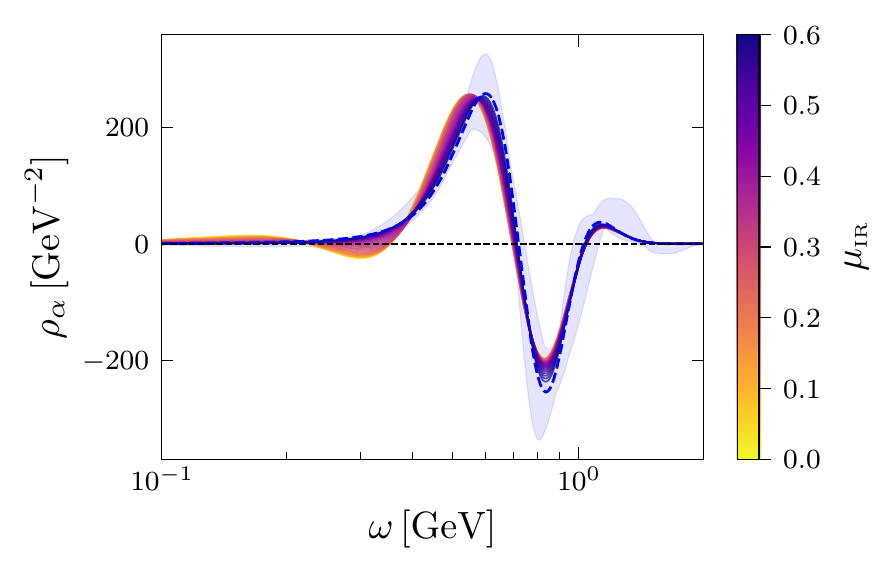}\label{fig:comp_IR}}%
    \subfloat[Scan of $\rho_\alpha$ along the flat $\mu_\UV$ direction]{%
		\includegraphics[width=.5\linewidth]{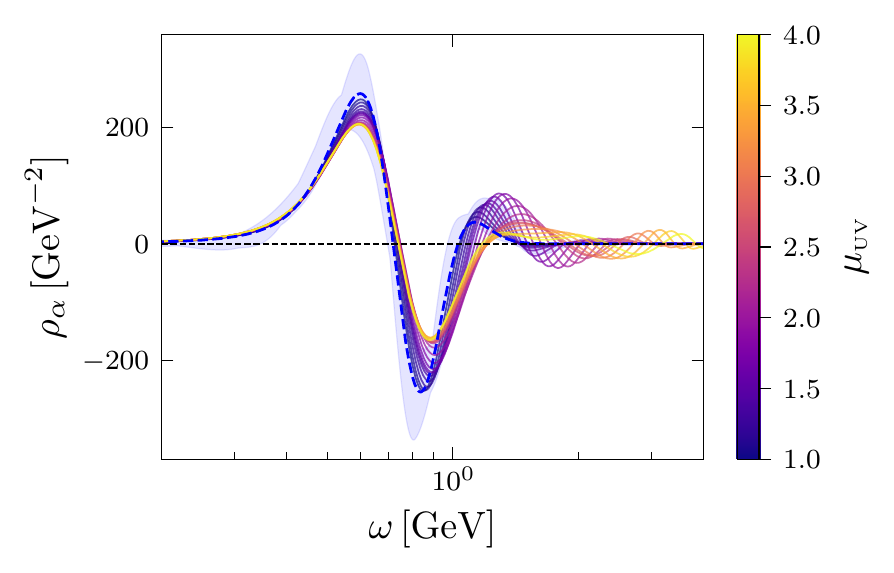}\label{fig:comp_UV}}%
    \caption{Behaviour of the spectral function when varying the midpoints $\mu_\IRUV$ of the transition kernels to the asymptotic IR (\Cref{fig:comp_IR}) and UV regimes (\Cref{fig:comp_UV}). The respective values of the parameters are colour-coded. The resulting scan of the spectral functions is compared to the final result with maximally enhanced asymptotics, displayed with a dashed blue line. The error band obtained by varying the parameters of the asymptotics---as indicated in \Cref{fig:parameterscan}---is given by the shaded blue area.}
	\label{fig:aysmptotics_comparison}
\end{figure*}

For the gluon and the ghost propagators, the leading order IR and UV asymptotics are known analytically; see~\cite{Cyrol:2018xeq} and references therein. In the infrared, the decoupling solution of the ghost is characterised by a constant propagator dressing $Z_c \equiv Z_c(p=0)$. On the other hand, the gluon propagator is dominated by the ghost loop polarisation diagram in the IR, since the gluon propagator itself has a mass gap and decouples in the infrared. This results in a $p^2\log p^2$ contribution in the IR regime; for a detailed discussion thereof, see~\cite{Cyrol:2018xeq}.

Using the definition of the strong coupling \labelcref{eq:prop_coupl}, we see that it has the same, but negative, IR behaviour as the inverse gluon dressing, up to a constant contribution from the ghost dressing. From \labelcref{eq:rho_alpha_IR}, we can then infer the asymptotic behaviour of the spectral function as
\begin{align}\label{eq:spec_asymp_IR}
    \rho_{\alpha, \IR}(\omega)\sim \omega^2
\,,
\end{align}
analogously to~\cite{Cyrol:2018xeq}. The UV asymptotic behaviour of the strong coupling is well known from perturbative calculations and reads
\begin{align}
    \alpha_{s, \UV}(p) \sim\frac{1}{\log(p^2)}
\,.
\end{align}
The asymptotic behaviour of the spectral function follows directly from \labelcref{eq:rho_alpha} and we obtain
\begin{align}\label{eq:spec_asymp_UV}
    \rho_{\alpha,\UV}(\omega) \sim -\frac{1}{\omega^2(\pi^2 + \log(\omega^2)^2)}
\,.
\end{align}

\section{Reconstruction details}

\subsection{GPR basics}\label{app:GPR}

Here, we briefly summarise the main aspects of the GPR reconstruction procedure. For a more detailed overview, we refer to earlier works~\cite{Horak:2021syv, Pawlowski:2022zhh}.

We assume our knowledge of the spectral function $\rho(\omega)$ before making observations of the correlator to be described by a GP prior, written as
\begin{align}
    \rho(\omega) \sim \mathcal{GP}(\mu(\omega), C(\omega, \omega))
\,,
\end{align}
where $\mu, C$ denote the mean and covariance. The conditional posterior distribution for $\rho(\omega)$ given observations of the propagator $G_i$ at $N_G$ discrete Euclidean frequencies $p_i \equiv [\boldsymbol{p}]_i$ can be derived in closed form,
\begin{align}
    \rho(\omega)|G(\boldsymbol{p}) \sim &\, \mathcal{GP}\Bigl(\boldsymbol{w}^T(\omega)(\boldsymbol{W} + \sigma_n^2 \mathbbm{1})^{-1}G(\boldsymbol{p}), \nonumber \\[1ex]
    & \hspace{-.5cm}k(\omega,\omega) - \boldsymbol{w}^T(\omega)(\boldsymbol{W} + \sigma_n^2 \mathbbm{1})^{-1}\boldsymbol{w}(\omega)\Bigr)
\,,
\end{align}
where
\begin{align}\label{eq:posterior}
    &[\boldsymbol{w}]_i(\omega) = \int \text{d}\omega'\, K(p_i, \omega') C(\omega', \omega)\,,\nonumber \\[1ex]
    &[\boldsymbol{W}]_{ij} = \int\text{d}\omega'\text{d}\omega'' \, K(p_i, \omega') K(p_j, \omega'') C(\omega', \omega'')
\,.
\end{align}
This is essentially equivalent to a standard result in probability theory for the closed-form expression of a conditional multivariate normal distribution, but defined with a continuum of random variables due to being a Gaussian \emph{process}, as well as additional applications of the integral transformation one seeks to invert. The equivalence becomes more concrete in practice when the GP is evaluated for a finite set of predictions; however, the choice of inference points $\omega$ is arbitrary within the given domain. In the above expressions, $\mu(\omega)$ has been set to zero, since a GP can be fully specified by its second-order statistics and the prior mean can be absorbed into $C$. The GP in \labelcref{eq:posterior} encodes our knowledge of the spectral function after making observations of the correlator and accounting for observational noise with variance $\sigma_n^2$.

The covariance $C(\omega,\omega')$ is commonly defined via a so-called kernel function with a small number of hyperparameters, which may be subject to optimisation based on the associated likelihood. A widely used parametrisation is the radial basis function (RBF) kernel, defined as
\begin{align}\label{eq:rbfkernel}
    C_\RBF(\omega,\omega') = \sigma_C^2\, \exp\left(-\frac{(\omega - \omega')^2}{2l^2}\right)
\,,
\end{align}
where the parameter $\sigma_C$ controls the overall magnitude and $l$ is a generic length scale.

\subsection{Incorporating asymptotic information}\label{app:kernel-asymp}

With the knowledge of the IR and UV asymptotics, cf.~\Cref{app:coupling-asymp}, an appropriate bias can be introduced.
It is chosen such that the kernel is restricted to the specified functional basis as described in \Cref{sec:GPR-reco}, while retaining the flexibility of the RBF kernel in the central region.

In order to achieve a smooth transition between the biased and unbiased kernels, we employ smooth step functions of the form
\begin{align}\label{eq:theta}
    \theta^{\pm}(\omega; \mu, \ell) = \frac{1}{1 + \exp(\pm 2(\omega-\mu)/\ell)}
\,.
\end{align}
The full kernel can then be written simply as a sum of the individual contributions,
\begin{align}\label{eq:asympkernel}
    k(\omega,\omega') = k_\RBF(\omega,\omega') + k_\IR(\omega,\omega') + k_\UV(\omega,\omega')
\,,
\end{align}
where
\begin{align}
    &k_\RBF(\omega,\omega') = \theta_\IR^+(\omega) \, \theta_\IR^+(\omega') \, \theta_\UV^-(\omega) \, \theta_\UV^-(\omega') \, C_\RBF(\omega, \omega') \nonumber \\[1ex]
    &k_\IR(\omega,\omega') = \theta_\IR^-(\omega) \, \theta_\IR^-(\omega') \, \rho_\IR(\omega) \, \rho_\IR(\omega') \nonumber \\[1ex]
    &k_\UV(\omega,\omega') = \theta_\UV^+(\omega) \, \theta_\UV^+(\omega') \, \rho_\UV(\omega) \, \rho_\UV(\omega') \nonumber
\,.
\end{align}
The midpoints of the transition functions $\theta_\IRUV$ are specified by $\mu_\IRUV$ and their steepness is controlled by $\ell_\IRUV$.

\begin{figure*}[t]
	\centering
	\subfloat[$\chi^2$ of the reconstruction under the variation of $\mu_\IRUV$.]{%
		\includegraphics[width=0.5\linewidth]{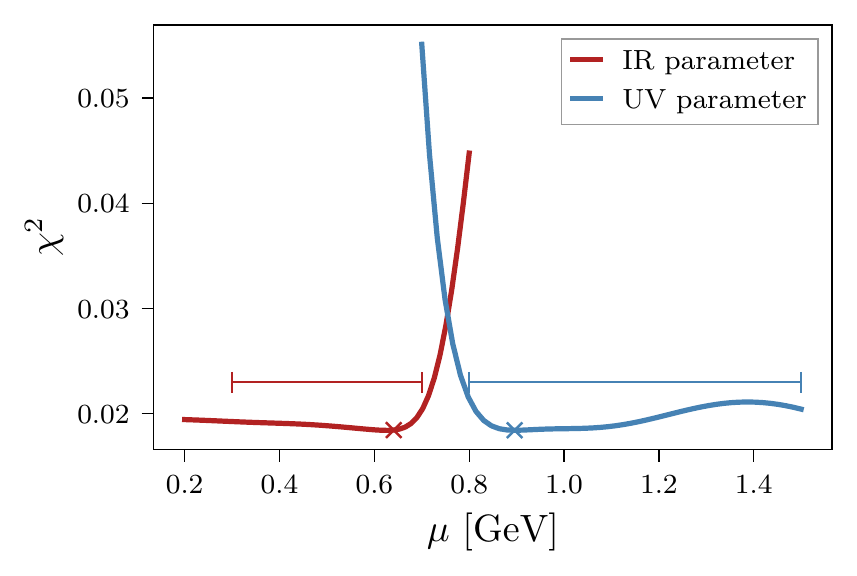}\label{fig:chi2mu}}%
	\subfloat[$\chi^2$ of the reconstruction under the variation of $\ell_\IRUV$.]{%
		\includegraphics[width=0.5\linewidth]{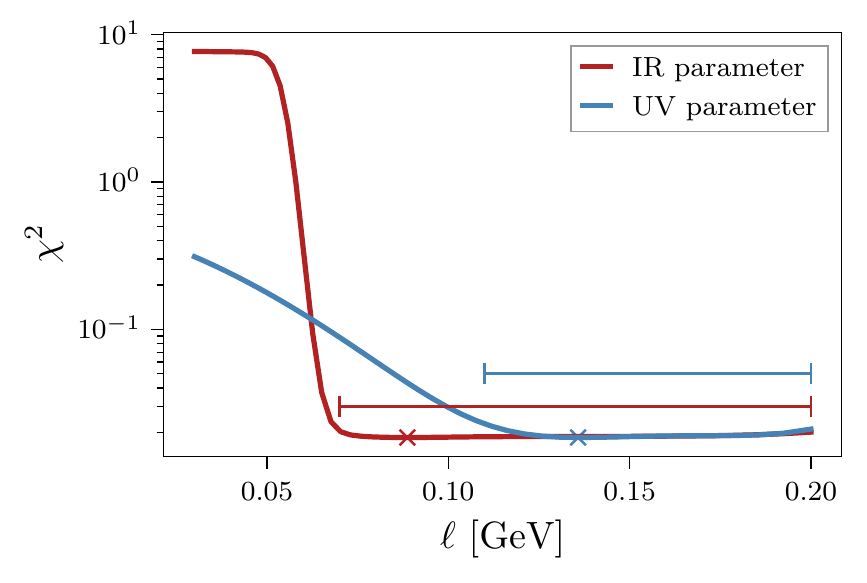}\label{fig:chi2l}}
	\caption{Scans of the bias parameters defined in \labelcref{eq:asympkernel}. We compare the quality of the dressing reconstruction---quantified by $\chi^2$---when varying the midpoint positions of the bias transition $\mu_\IRUV$ (\Cref{fig:chi2mu}) as well as its steepness $\ell_\IRUV$ (\Cref{fig:chi2l}). The values of the bias parameters chosen for the reconstruction are marked by crosses. This choice maximises the size of the regions dominated by the coupling infrared and ultraviolet asymptotics, while producing small $\chi^2$ reconstructions of the data. Additionally, the parameters are then scanned in the flat directions, indicated by the horizontal bars, in order to obtain the error estimation for the reconstruction results shown in \Cref{fig:spec_func_rep_coupling}.}
	\label{fig:parameterscan}
\end{figure*}

\subsection{GP kernel hyperparameters}\label{app:hyperparams}

Since the hyperparameters of the GP kernel control the behaviour of the resulting spectral function, their choice is a pivotal step in the reconstruction. They are commonly determined via numerical optimisation by minimising (conventionally) the negative log-likelihood (NLL),
\begin{align}\label{eq:NLL}
	-\log p(G(\boldsymbol{p})|\boldsymbol{\sigma}) = \frac{1}{2} G(\boldsymbol{p})^T \left(\boldsymbol{W_\sigma} + \sigma_n^2 1 \right)^{-1} G(\boldsymbol{p}) + \nonumber \\[1ex] 
	\frac{1}{2}\log \det(\boldsymbol{W_\sigma} + \sigma_n^2 1) + \frac{N}{2} \log 2\pi
\,,
\end{align}
where the dependence on the kernel hyperparameters $\boldsymbol{\sigma}$ is emphasised by an index.

The number of hyperparameters increases significantly when including the bias term that enforces the correct asymptotics \labelcref{eq:asympkernel}. Hence, the two parameters of the bare RBF kernel are chosen first by minimising \labelcref{eq:NLL}. The asymptotics are then introduced in the far IR/UV and shifted towards the center, all while monitoring the quality of the interpolation of the dressing by computing $\chi^2$ at each step. We compare $\chi^2$ instead of the NLL for different bias parameters, since the second term in \labelcref{eq:NLL} constitutes a complexity penalty term. When considering an explicit functional basis, such a term is inherently in opposition to the constraint for the analytically known asymptotics and is therefore excluded. We observe that the bias kernel parameters have an open direction towards vanishing bias, e.g., for small $\mu_\IR$ and large $\mu_\UV$. Spectral functions with $\mu_\UV > \SI{1.5}{\giga\eV}$ show a growing number of smaller oscillations in the UV, which are a remnant of the global length scale introduced in the RBF kernel; see \Cref{fig:comp_UV}. Accordingly, models in this parameter region can be ruled out as sensible descriptions of the underlying physics of the coupling. For $\mu_\IR < \SI{0.25}{\giga\eV}$, the resulting spectral functions do not change significantly; see \Cref{fig:comp_IR}. The change of $\chi^2$ when varying the asymptotic kernel parameters is shown in \Cref{fig:parameterscan}, with the final settings used for the reconstruction indicated by crosses. These parameters are explicitly chosen to maximise the regions dominated by the asymptotics, without significantly increasing the error of the coupling reconstruction. All final parameters of the GP model used to compute the results reported in this work are listed in \Cref{tab:hyperparams}. 

\begin{table}[b]
	\begin{center}
	\begingroup
    \renewcommand*{\arraystretch}{1.5}
		\begin{tabular}{l  c  c  c  c  c  c }
			Parameter & $\sigma_\RBF$ & $l_\RBF$ & $\mu_\UV$& $\ell_\UV$ & $\mu_\IR$ & $\ell_\IR$ \\
			\hline
			Value & ~67.399~ & ~0.074~ & ~0.890~ & ~0.137~ & ~0.637~ & ~0.090~ \\
		\end{tabular}
		\endgroup
    \caption{Hyperparameters for the combined RBF and fixed-asymptotics kernel, as defined in \labelcref{eq:asympkernel,eq:rbfkernel}.}
	\label{tab:hyperparams}
	\end{center}
\end{table}

The error estimation for the spectral function via the covariance of the posterior distribution does not include the systematic error that arises from the choice of the model, in particular regarding different values of the hyperparameters. However, we observe that enforcing the maximally large asymptotic regimes leads to the predicted posterior covariance being comparatively small as the model is now highly restricted; see \Cref{fig:gluon-spectral}. Hence, the error is estimated by varying the bias parameters in a region where $\chi^2$ is small, but the effect of different parameter choices is non-negligible, while unphysical oscillations remain largely suppressed. This region is marked in \Cref{fig:parameterscan} by horizontal bars. When considering $\mu_\UV$ larger than indicated in this region, a substantial amount of oscillations is introduced in the spectral function as mentioned above. For $\mu_\IR$ smaller than indicated in \Cref{fig:chi2mu}, the spectral function vanishes in the IR. However, it can then differ from the expected $\omega^2$ behaviour. The largest variations in the resulting spectral functions under these changes of the hyperparameters are then used as error estimates for the reconstruction results, shown in \Cref{fig:spec_func_rep_coupling}. Since the deviations of the predictions at the edges of the parameter space tend to be maximised in particular regions for certain parameter combinations, e.g., when $\mu_\UV$ and $\ell_\UV$ are both small, the error band shows a few distinct kinks.

\bibliography{main}

\end{document}